\documentclass[journal=jacsat,manuscript=article]{achemso}

\usepackage[version=3]{mhchem}

\usepackage{graphicx}  % needed for figures
\usepackage{dcolumn}   % needed for some tables
\usepackage{bm}        % for math
\usepackage{amssymb}   % for math
\usepackage{braket}
\usepackage{amsmath}

\usepackage{wrapfig}

\title{Quantum wires and waveguides formed in graphene by strain}

\author{Y.~Wu} \affiliation{Department of Physics and Astronomy, University of California, Riverside, California 92521, USA}
\author{D.~Zhai} \affiliation{Department of Physics and Astronomy, Ohio University, Athens, Ohio 45701-2979, USA}
\author{C.~Pan} \affiliation{Department of Physics and Astronomy, University of California, Riverside, California 92521, USA}
\author{B.~Cheng} \affiliation{Department of Physics and Astronomy, University of California, Riverside, California 92521, USA}
\author{T.~Taniguchi} \affiliation{Advanced Materials Laboratory, National Institute for Materials Science, Tsukuba, Ibaraki 305-0044, Japan}
\author{K.~Watanabe} \affiliation{Advanced Materials Laboratory, National Institute for Materials Science, Tsukuba, Ibaraki 305-0044, Japan}
\author{N.~Sandler} \affiliation{Department of Physics and Astronomy, Ohio University, Athens, Ohio 45701-2979, USA}\email{sandler@ohio.edu}
\author{M.~Bockrath} \affiliation{Department of Physics, The Ohio State University, Columbus, OH 43210, USA}
\email{bockrath.31@osu.edu}

\begin{document}

\begin{abstract}
Confinement of electrons in graphene to make devices has proven to be a challenging task. Electrostatic methods fail because of Klein tunneling, while etching into nanoribbons requires extreme control of edge terminations, and bottom-up approaches are limited in size to a few nm. Fortunately, its mechanical flexibility raises the possibility of using strain to alter graphene's properties and create novel straintronic devices. Here, we report transport studies of nanowires created by linearly-shaped strained regions resulting from individual folds formed by layer transfer onto hexagonal boron nitride. Conductance measurements across the folds reveal Coulomb blockade signatures, indicating confined charges within these structures, which act as quantum dots. Along folds, we observe sharp features in traverse resistivity measurements, attributed to an amplification of the dot conductance modulations by a resistance bridge incorporating the device. Our data indicates ballistic transport up to $\sim$1 $\mu$m along the folds. Calculations using the Dirac model including strain are consistent with measured bound state energies and predict the existence of valley-polarized currents. Our results show that graphene folds can act as straintronic quantum wires.

\end{abstract}

\vspace{.15 in} 

Graphene, the one-atom-thick layer of carbon atoms arranged in a honeycomb structure, has remarkable electronic and mechanical properties\cite{geim2007rise,neto2009electronic}. Electric fields\cite{castro2007biased,mak2009observation,ohta2006controlling}, spatial confinement\cite{Han2007energy,son2006half,Li2008,jacobse_electronic_2017,baringhaus_exceptional_2014} and periodic potentials\cite{dean2013hofstadter,ponomarenko2013cloning,cheng2016gate,wang2015topological,hunt2013massive} are just some of the methods used to manipulate its electronic behavior. As an atomically thin membrane, graphene is highly flexible\cite{eda2008large,lee2008measurement} and strain engineering can also be used to control its electronic properties\cite{pereira2009strain,levy2010strain,guinea2010energy,Vozmediano2010,Carrillo-Bastos2014,Carrillo-Bastos2016, rasmussen_electronic_2013}. It was reported, for example, that in graphene nanobubbles strain-induced pseudomagnetic fields can reach magnitudes greater than 300 T\cite{levy2010strain} while creating strong density of state modulations. Wrinkled or rippled graphene on SiO$_2$ substrates\cite{Ni2012, Zhu2012, Shioya2015, willke_local_2016} and suspended samples\cite{Bao2009} were also carefully studied, including investigation of electron-phonon scattering\cite{Ni2012}, Kelvin probe microscopy of fold resistivity,\cite{willke_local_2016} potential fluctuations\cite{Shioya2015}, and the average impact on electrical resistance across many samples with large number of folds\cite{Zhu2012}. Furthermore, scanning tunneling microscopy measurements\cite{klimov2012electromechanical} demonstrated that pseudomagnetic fields confine electrons in deformed regions locally creating quantum dots. Despite this progress, the use of strain to produce building blocks for electronic systems, such as quantum wires and coherent electron waveguides, although promising\cite{andrei2017folds} remains elusive, while even basic quantum transport properties are still unexplored. 

In this work, we present transport measurements on graphene devices with well defined, isolated, linearly-shaped strained regions created by nanoscale-width folds. These folds are produced by placing graphene membranes onto hexagonal boron nitride (hBN) substrates\cite{dean2010boron}, which are supported by oxidized Si wafers. Attached source and drain electrodes enable measuring the differential conductance $dI/dV$ while tuning the charge density in the graphene using the Si wafer as a gate electrode. Charge transport is measured in perpendicular and parallel directions with  respect to the fold axis orientation. Remarkably, measurements carried out with a perpendicular incident current exhibit single-electron charging and Coulomb blockade features, clear signs of charge confinement. Observed quantum level spacings and charging energies reveal the confinement to occur in regions of dimensions of the strained-fold width, indicating that the fold acts as a finite length one-dimensional (1D) quantum wire, likely broken into several shorter conducting segments by disorder. Experimental observations are in agreement with theoretical calculations that model the behavior in terms of a strain-induced pseudomagnetic field \cite{pereira2009strain,Carrillo-Bastos2016} which creates barriers confining electrons to nanoscale channels. Transport with parallel incident current reveals sharp structures reminiscent of Fano-like features\cite{Goeres2000}, suggesting interference between conduction paths entirely within the graphene sheet, and states bound in the deformed region. 

In addition to confining electrons to one dimension, the pseudomagnetic field is predicted to filter electrons from individual valleys\cite{Carrillo-Bastos2016}. We show the filtering depends on the incident angle of the incoming current, a unique feature that will enable the realization of valleytronic devices\cite{garcia2008fully,Akhmerov2008,Rycerz2007,Carrillo-Bastos2016} by appropriate design of fold geometries. As the incidence angle can be controlled experimentally, a linearly strained graphene device such as the one reported in this work, can potentially serve as a valley filter for valleytronics applications.

%%%Figure1

Figure~1a-c shows the typical device geometry. Devices are fabricated using Elvacite polymer layers and dry transfer to encapsulate graphene between hBN layers\cite{wang2013one}. Figure~1a shows a Scanning Electron Microscope (SEM) image of a graphene flake with a linear strain (dark line) in the center after its dry-transfer to hBN. The cyan dashed outline encloses the graphene area later made into a device via etching. We distinguish two types of isolated folds (Supporting Information section 1)\cite{Zhu2012}. Type I is characterized by approximately Gaussian shapes with heights -measured by atomic force microscopy- between $10$ to $20$ nm, while Type II exhibits a flat-plateau geometry with reduced heights between $2$ to $5$ nm. It is likely that Type II folds result from Type I bent laterally to contact the graphene sheet\cite{Zhu2012}. Following fold identification, a top BN layer is added to encapsulate the graphene (Fig.~1b). Four electrical contacts are fabricated by an edge contact technique\cite{wang2013one} on the four sides of the BN/graphene/BN sandwich structure (Fig.~1b inset). A colorized SEM image of a device is shown in Fig.~1c, consisting of the encapsulated graphene (blue region) and electrical contacts (yellow regions) with a red dashed line indicating the location of the linear strain. Graphene sheet contacts are labeled G1 and G2, while nanowire contacts (as well as the top sides of the sheet) are labeled N1 and N2. The linear strain is located at the device center. 

%%%Figure2

To study the electronic transport properties of the system, we first measure $dI/dV$ using the setup shown in Fig.~1c, with perpendicular current flow, i.e. between contacts G1 and G2, while floating contacts N1 and N2. The experiment is performed using a lock-in amplifier with a source-drain dc voltage ($V_{sd}$) applied to G1 in addition to a small ac bias ($dV$) voltage. The ac current ($dI$) is measured vs. $V_{sd}$ and gate voltage $V_{g}$ applied to the Si back gate. Figure~2a shows a $dI/dV$ color plot  vs. $V_{sd}$ and $V_{g}$ at $T = 1.5$ K. Remarkably, the data show Coulomb diamond features similar to those previously observed in nanotubes\cite{laird_quantum_2015} or confined graphene\cite{ponomarenko2008chaotic}, a phenomena characteristic of charge confinement that occurs in quantum dot (QD) geometries. The bright lines (large values of $dI/dV$)  represent the opening of seemingly quantized transport channels across the system. We estimate the charging energy $E_{c}$ for these regions from the voltage scale of the diamonds\cite{Kouwenhoven1997} and obtain $E_c\sim 20\pm 8$ meV for the diamonds marked by the white dashed lines. Assuming the capacitance scales with the fold length (similar to carbon nanotubes) we obtain effective lengths of approximately $L \sim 250$ nm. This suggests that disorder, lack of uniformity, or change of orientation (e.g., a kink) along the fold creates an effective shorter length (as compared with the distance N1-N2 $\sim 1.5$ $\mu$m) that dominates the Coulomb blockade (CB) regime at zero bias. 

Figure~2b shows a $V_g$ line trace at $V_{sd} = 0$ V (along the red dashed line in Fig.~2a). CB peaks are observed, separated by low conductance regions. These peaks occur when neighboring charge states become degenerate, enabling current flow by single-electron charging and discharging cycles. The low conductance regions, which appear as the dark blue diamond areas in Fig.~2a, correspond to a fixed charge state on the fold, and indicate that contacts N1 and N2 only make good electrical contact to the graphene on one side of the nanowire. The Fig.~2b inset shows the corresponding $dI/dV$ versus $V_g$ at different temperatures for the peak enclosed by the red rectangle in main panel. The curves shift slightly as the temperature changes, and $V_g$ is shifted accordingly to maintain the peak alignment. The amplitudes of the differential conductance peaks become smaller, and their widths become larger with increasing temperature. This indicates a resonant single quantum level transport process\cite{Kouwenhoven1997}. Coulomb oscillations persist up to the largest temperature measured of $T= 20$K, with little temperarure dependence in the conductance minima, as expected based on the measured charging energy $E_{c} \sim 20$ meV.

A zoom-in image of data taken at finite bias from inside the magneta rectangle of Fig.~2a is shown in Fig.~2c. Parallel lines observed outside the insulating diamond-shaped regions correspond to tunneling through discrete energy excitations (marked as arrows in the figure). 

Figure~2d shows a line trace of $dI/dV$ versus $V_{sd}$ at $V_{g}=25$ V (along the white dashed line in Fig.~2c). Six peaks are evident in the figure. The left five peaks correspond to discrete energy excitations marked with arrows in Fig.~2c, while the right peak corresponds to reaching the condition where current flow is first enabled at the CB diamond boundary. Analysis of the data \cite{Kouwenhoven1997} reveals uniform energy level spacings of $\Delta E\sim 2$ meV. Assuming a finite length 1D quantum wire geometry model without broken valley degeneracy (relaxing this assumption leads to similar results), the energy level spacing can be estimated by $\Delta E\sim h v_F/(2L')$, where $v_F\sim 10^6$ m/s is the Fermi velocity of the confined states and $L'$ is an effective length $\sim 1.0$ $\mu$m, a value larger than the scale $L$ estimated by the charging energy but similar in order of magnitude (see Supporting Information section 2). This indicates a ballistic mean free path along the nanowire up to  $\sim 1$ $\mu$m. Interpreting the data as arising from transport through dots of different sizes as suggested by the inferred lengths $L$ and $L'$ and the several downward slopes in Fig.~2c, (see e.g. Abulizi et al. \cite{abulizi_full_2016}) is therefore consistent with a distorted or disordered fold extended along the sample between contacts N1 and N2.

We now turn to measurements with parallel incident current. Fig.~2e shows the conductance versus $V_g$ that is finite for the voltage range explored with no Coulomb peaks observed for this flow direction. This is consistent with the large resistance ratio between the quantum wire and flat sheet regions so that the bulk of the current flows through the flat regions. The fold is detected however, by its effects on the transverse resistance $R_{xy}=V_{xy}/I$ versus $V_{g}$ (where $I$ is the current in the direction parallel to the fold). The measurement setup is shown in the right inset of Fig.~2f, with current passed through contacts N1 and N2 and voltage $V_{xy}$ measured across G1 and G2. Transverse resistance data taken at $T = 1.5$ K shows small oscillations superimposed on large jumps that can be attributed to contributions from the confined states in the fold region and extended states in the graphene layer. The increased sensitivity of the transverse measurement to the presence of the finite size wire can be explained by the graphene sheets and voltage probes forming a Wheatstone bridge geometry with the nanowire dot acting as the bridge resistance (see Supplemental Information section 3).  The sharp adjacent dip and peak in the data around $V_g=-30 V$ is reminiscent of Fano interference processes\cite{fano1961effects}, which may occur between states in graphene and those dwelling temporarily on the nanowire, however more work will be required to fully elucidate the transport mechanism leading to this phenomenon.

%%%%%%%%Figure3

To further explore the  nature of the confinement, we measured several different samples within the same device setup as Fig.~1c. The tunnel barriers between the linearly strained regions and the flat graphene layer and/or contacts N1 and N2 were found to vary between highly reflecting (yielding CB diamonds), and highly transmitting [yielding Fabry-Perot resonances~\cite{liang2001fabry}]. The measurement setup is schematically shown in  Fig.~3a. As before, data is obtained with the differential conductance $dI/dV$ taken across the linear strain with floating contacts N1 and N2 corresponding to open switches in Fig.~3a. Fig.~3b shows data taken for one Fabry-Perot type device, where neighboring diagonal linear features (dashed lines) appear with similar slopes over the entire gate voltage range explored. These features arise when the electrochemical potentials of the source and drain electrodes align with Fabry-Perot resonances in the finite size wire. For comparison, a set of data taken across the linear strain with electrodes N1 and N2 grounded is shown in Fig.~3c (Fig.~3a with closed switches), as a color plot for $dI/dV$ as a function of $V_{sd}$ and $V_{g}$. The criss-cross pattern associated with the Fabry-Perot interference disappears. Instead, discrete vertical linear features (dashed line) in the spectra can be identified. 

This behavior can be accounted for by the schematic equivalent circuit diagram shown in Fig.~3d. In this setup, the nanowire appears as a high-impedance tunnel barrier, along with resistors $R_1$ and $R_2$, and the switch $S_1$. $R_1$ represents the flat graphene resistance connected to the nanowire on one side, while $R_2$ corresponds to the resistance to ground on the same side of the nanowire on the sheet. (A similar resistor network could be shown for the other side of the device, but is omitted for simplicity.) When N1 and N2 are floating, the switch $S_1$ in Fig.~3d is open (contact position b), only a negligible voltage drop occurs across $R_1$ and essentially the full source-drain voltage appears across the nanowire. On the other hand, when N1 and N2 are grounded, the switch $S_1$ is closed, (contact position c), the voltage is divided by the voltage division circuit formed by $R_1$ and $R_2$.  The vertical slope of the features shows that the electrons crossing the device through the nanowire have an energy near the electrochemical potential of the ground electrode, indicating that $R_2<<R_1$,  consistent with the region of the charge confinement occurring near the nanowire. Based on this, we expect that the criss-cross pattern would be visible at larger voltages, scaled by $\sim R_1/R_2$, however to avoid potential electronic damage to the device, such larger voltages were not applied.

Taken together, the observed quantum levels and Fabry-Perot resonances show that the folds act as quantum wires and waveguides. While the structure of the folds is unknown after the capping BN layer is added, we confirmed that the initial folds are necessary to observe the electron confinement behavior. The stress associated with being placed between two layers with van der Waals adhesion forces could distort the initial fold shape, for example changing a type I structure to a type II, and Coulomb oscillations persist enhancing the strain and maximum curvature of the folds (see Supplemental Information, section 4). To gain insight into the origin of these features, we consider a model with a strained out-of-plane deformation depicted in Fig.~4a, which should be appropriate for modeling any sharp crease in graphene.  We consider the in-plane displacement (expected from the mismatch with the hBN lattice) to be negligible as compared to the out of plane displacement. Due to the strong in-plane boding of carbons it is expected that the strain does not produce lattice defects\cite{kim_multiply_2011}. Nevertheless, it is well established that strain strongly modifies the local density of states, giving rise to localized states in deformed regions\cite{pereira2009strain,Vozmediano2010,guinea2010energy,levy2010strain,klimov2012electromechanical,Carrillo-Bastos2014}. At low energies, the electron dynamics are described via a massless Dirac equation, with strain producing an effective pseudo-vector potential. In the valley isotropic basis $(\psi_{KA}, \psi_{KB}, -\psi_{K'B}, \psi_{K'A})$, ($A, B$ denote the two-atom basis), the Hamiltonian is\cite{pereira2009strain,Vozmediano2010,guinea2010energy,Carrillo-Bastos2016}:
$H_{K,K'}=v_F\vec{\sigma}\cdot\left(\vec{p}\mp\vec{A}\right)$. Here,
$K,K'$ labels the two valleys, $\vec\sigma$ is the Pauli matrix vector, and $\vec{p}$ the momentum. The strain tensor components $\epsilon_{ij}=\frac{1}{2}(\partial_j u_i+\partial_i u_j+\partial_i h \partial_j h)$, where $u_{i,j}$ and $h$ are in-plane and out-of-plane displacements, respectively and determine the pseudo-vector potential:
\begin{equation}
\vec{A}=\left(A_x,A_y\right)^T=\frac{\hbar\beta}{2a}\left(\epsilon_{xx}-\epsilon_{yy},-2\epsilon_{xy}\right)^T,
\end{equation}
where $a\approx1.42$\AA, $\beta\approx 3$, predicted within a $\sim 20\%$ range of variation\cite{Midtvedt2016}, and $\hat{x}$ ($\hat{y}$) is along the zigzag (armchair) direction\cite{Vozmediano2010}.
A Gaussian-shaped fold is represented by $h(y)=h_0e^{-y^2/b^2}$, with height $h_0$ and width $b$ as shown in Fig.~4a. The corresponding pseudo-vector potential is
\begin{equation}
A_{x}(y)=-\frac{\overline{\beta}\eta^2}{v_F}g\left(\frac{y}{b}\right),
\end{equation}
where $\overline{\beta}=\frac{\hbar\beta v_F}{2a}\approx7$eV, $\eta=\frac{h_0}{b}$, and $g(z)=2z^2e^{-2z^2}$. Fig.~4b shows the inhomogeneous amplitude of the pseudo-vector potential (blue) and its corresponding pseudomagnetic field $\vec{B} = \bigtriangledown \times \vec{A}$ (red).  

Scattered and bound states are obtained with scattering matrix  methods applied to incident states with energy $E$ and wave vector $\vec{k}$ related by: $\hbar v_F k_x=E\cos\theta$, $\hbar v_F k_y=E\sin\theta$, where $\theta$ is measured with respect to the zigzag crystalline direction. Inside the folded region $k_x\rightarrow k_x\mp A_x(y)$ and $k_y\rightarrow q_y=\sqrt{E^2-(\hbar v_F k_x\mp v_FA_x(y))^2}/\hbar v_F$. Thus, for appropriate values of $E, k_x$ and $A_x$, $q_y$ can become imaginary and render zero transmission.

Fig.~4c shows the transmission spectra through a fold along the zigzag direction for valley $K$ (results for $K'$ are mirror symmetric and not shown). Inside the original Dirac cone, the bright colored region represents non-zero transmission
scattering states that fades as $q_y\rightarrow 0$ and becomes zero for imaginary $q_y$ (blue region). At low energies (E $\lesssim 0.3$ eV), transmission is almost forbidden for all incident angles, indicating a highly reflecting fold (see Fig. 4c inset). For $E=-\hbar v_F k_x$ ($\theta=180^\circ$), sets of split resonances are visible. These states connect to the red dotted lines that represent the dispersion of bound states that propagate along the fold. 

The orientation of the fold with respect to the crystalline axis strongly affects the strength of the pseudomagnetic field, and hence the system transport properties. For a fold along an arbitrary angle $\alpha$ with respect to the zigzag direction, $A_{x}\rightarrow A_{x}\cos(3\alpha)$ ($\hat{x}$ along the fold axis), making a fold with axis close to armchair directions highly transmitting. Fig.~4d shows transmission and bound state energies for a fold with axis $1^\circ$ away from the armchair direction. Strong transmission occurs practically for all states (see Fig. 4d inset), with bound states confined to a narrow energy range. This transmission spectra is consistent with Fabry-Perot signatures observed in some experimental samples.

In addition to confinement, $\vec{A}$ induces valley polarization\cite{Carrillo-Bastos2014,Carrillo-Bastos2016,SettnesValleyFilter} as seen in Figs.~4e-f for $K$  and $K'$ respectively, as a function of $\theta$. For this fold orientation, it is achieved at incident angles greater (lesser) than $\pi/2$, being optimal for incidences close to armchair directions. 

These features are intuitively understood with a double-barrier vector potential model: the barrier height is set by $A_x$ (tuned through $\eta^2$) and determines the cutoff energy below which transmission via scattering states vanishes. The width $b$ determines energy and splittings of resonances. 

These results show that folds are capable of confining electrons and acting as nanowires. While this picture is directly applicable to a type I fold, type II fold can be thought of consisting of multiple nanowires in series and/or parallel (Supporting Information section 4), however a direct comparison to experiment is not possible as the precise fold structure following BN encapsulation is unknown.   
Although the data shown in this Report was obtained in monolayer graphene devices, similar behavior (Coulomb blockade and/or Fabry-Perot interference) is observed in bilayer and trilayer devices (data from bilayer device and additional data from monolayer devices shown in Supporting Information section 5).

%%%%Summary

In summary, transport properties of strained folds in graphene exhibit a rich behavior ranging from Coulomb blockade to Fabry-Perot oscillations for different fold orientations. Those exhibiting strong confinement, behave as electronic waveguides in the direction parallel to the fold axis, providing a new way to realize 1D conducting channels in two-dimensional graphene by strain engineering. Moreover, these geometries are promising candidates for the design of valley filter devices in current experimental settings.
%\bibliographystyle{unsrt}
%\clearpage

\bibliography{bibfile}

\noindent \textbf{Acknowledgements:}
We acknowledge discussions with D. Faria. This work was supported by DOE ER 46940-DE-SC0010597 (Y. W., C. P. , B. C. and M. B.) and NSF-DMR 1508325 (D.Z. and N.S.). Additional support for device fabrication was from the UCR CONSEPT Center.

\noindent \textbf{Supporting Information:} Additional discussion and figures concerning the structure of the folds and data from a fold in a graphene bilayer.

%\large{\textbf{References}

%\bibliography{largeD}{}
%\begin{thebibliography}{}
%\end{thebibliography}

\newpage

\begin{figure}
\begin{minipage}{\textwidth}
\begin{center}
\includegraphics[width=6in]{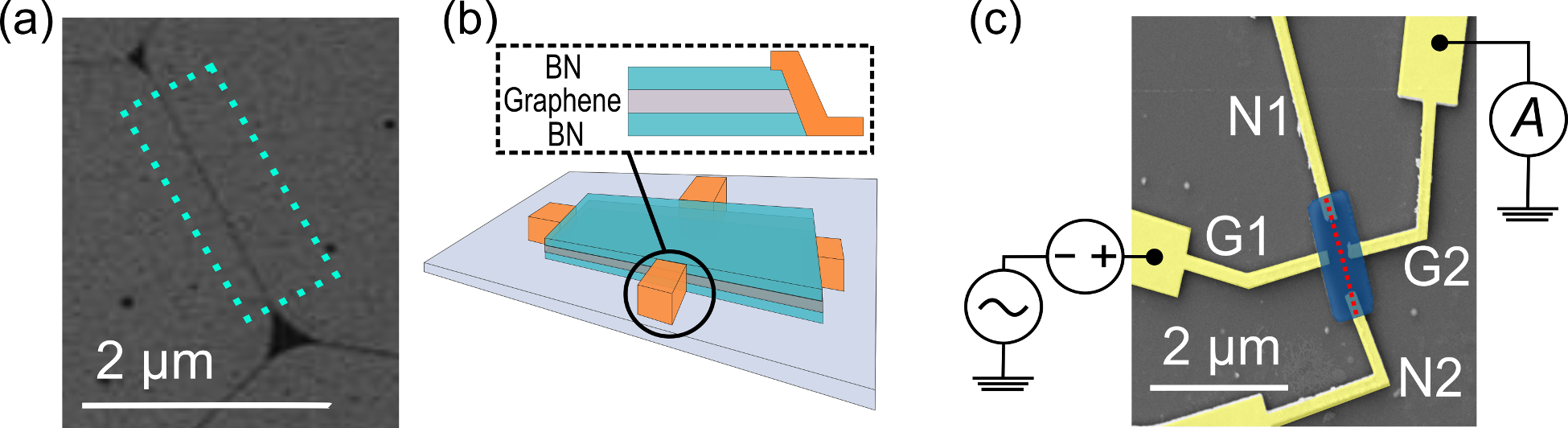}
\end{center}
\caption{Device geometry and experimental setup. ({a}) SEM image of graphene with a linear strain region due to a fold (dark line) in the center (cyan rectangle). ({b}) Device geometry. After etching to leave the region within the cyan rectangle in ({a}), the graphene is encapsulated with hBN. Contacts are made on each of the four sides. Top inset: details of the edge-contact geometry. ({c}) Colorized SEM image of device. Encapsulated graphene (blue) and the four contacts (yellow), respectively. Red dashed line: linearly-shaped strain region.}
\label{1}
\end{minipage}
\end{figure}

\newpage
\begin{figure}
\begin{center}
\includegraphics[width=6.5in]{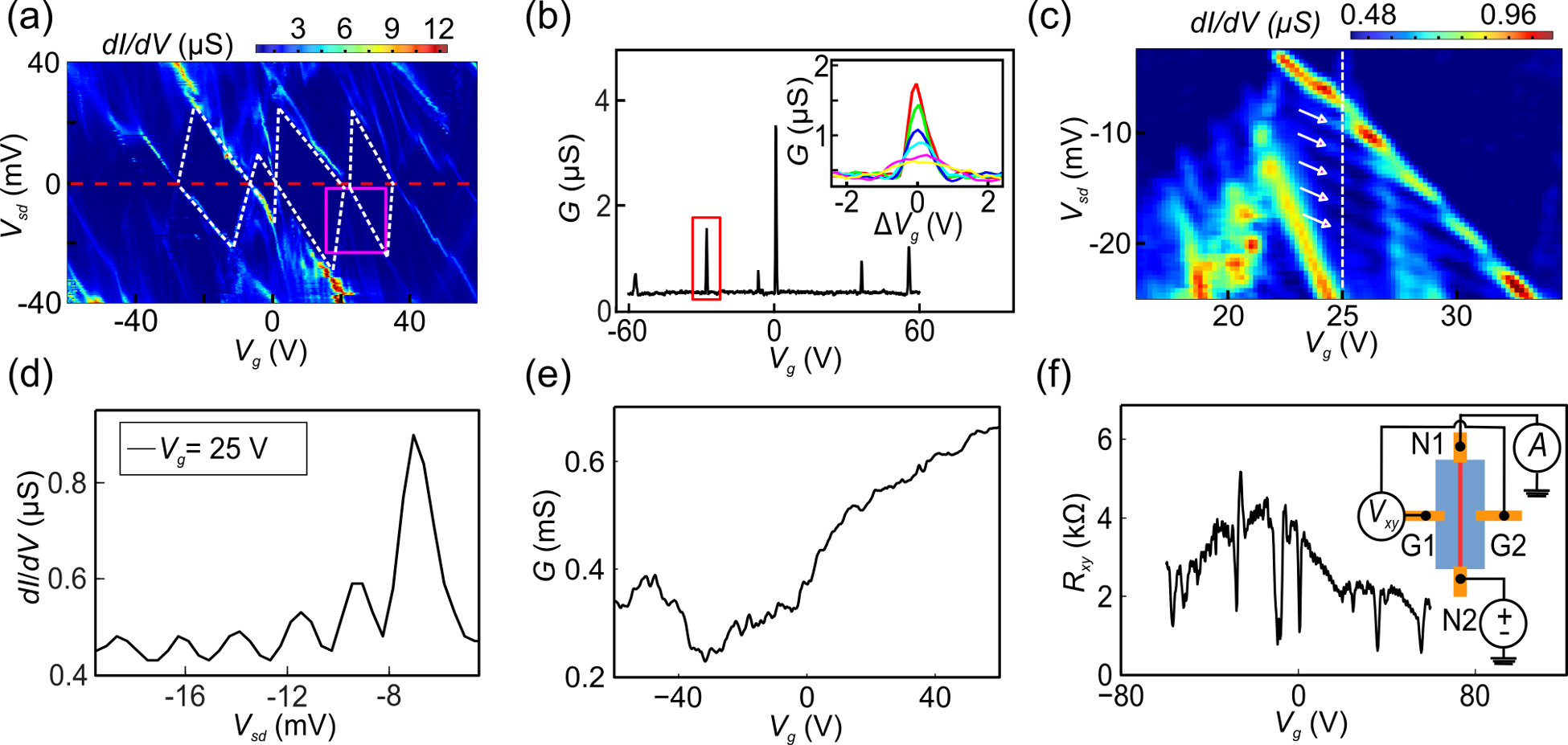}
\end{center}
\caption{{Low temperature transport data.} ({a}) Differential conductance vs. $V_{sd}$ and $V_{g}$ showing Coulomb diamond features (white dashed lines). ({b}) Line cuts taken at $V_{sd}=0$ V (red dashed line in {a}) Inset: differential conductance vs. relative gate voltage at different temperatures inside the red rectangular region. For each curve temperatures are: yellow, 20 K, pink 15 K, cyan 10 K, blue 6 K, green 4 K, red 2 K. ({c}) Zoomed-in figure of white rectangle region in {a}. ({d}) A line trace at $V_{g}=25$ V (pink dashed line) in {c}. ({e}) Transport with current parallel to fold showing conductance between contacts N1 and N2 vs. $V_g$. ({f}) $R_{xy}$ as a function of gate voltage. The inset shows the measurement setup, scale bar is $2$ $\mu$m. Red dashed line: location of the linear strain.}
\label{2}

\end{figure}

\begin{figure}
\begin{center}
\includegraphics[width=4.6in]{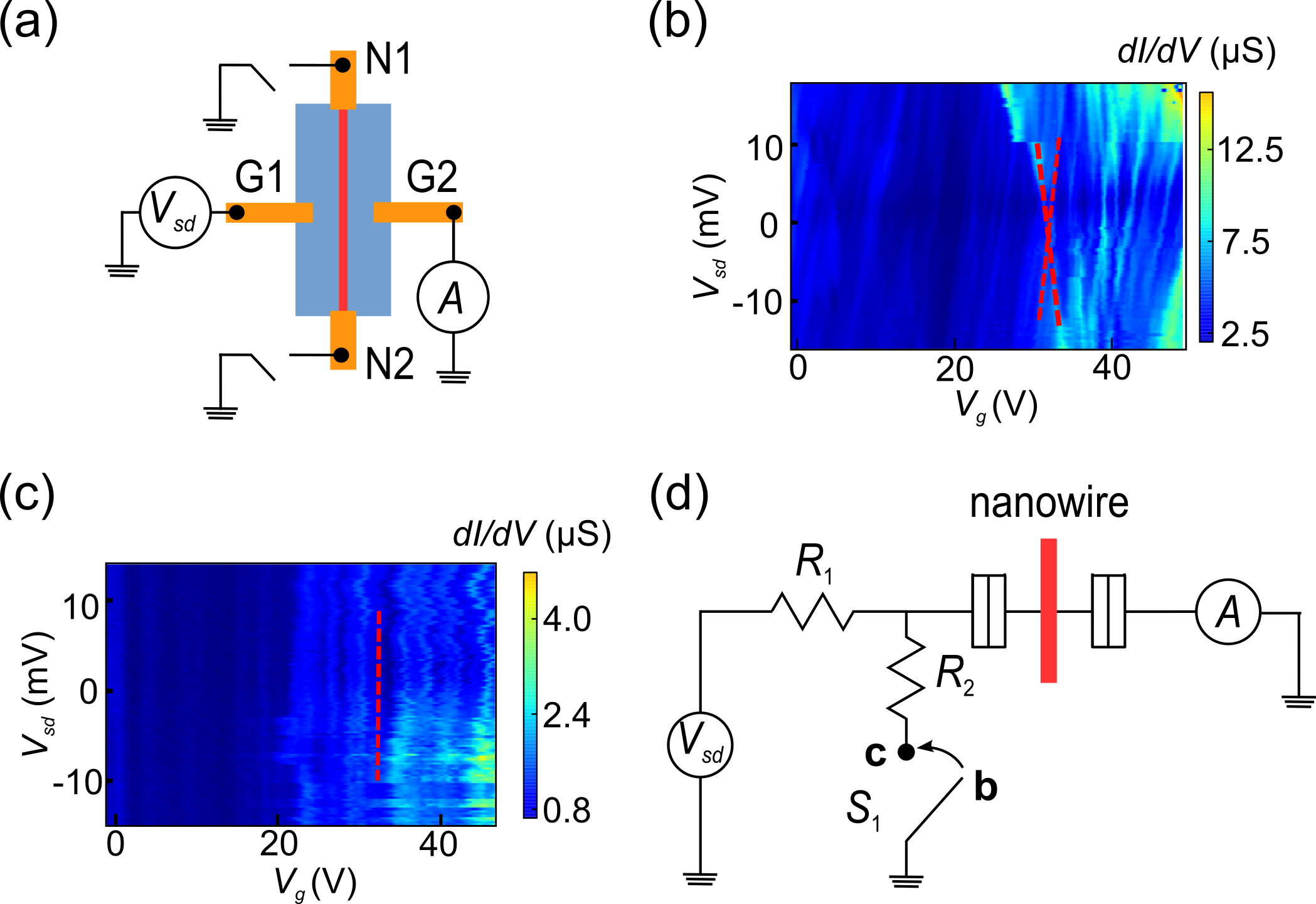}
\end{center}
\caption{{ Additional transport measurements on a device in the Fabry-Perot regime.}  ({a}) Measurement setup: differential conductance is measured across the fold (red line) with floating contacts N1 and N2.  ({b}) $dI/dV$ across the linear strain vs. $V_{sd}$ and $V_g$ of a Fabry-Perot type device. ({c}) The same color plot taken with a measurement set up shown in {a} but with grounded contacts N1 and N2. ({d}) Equivalent circuit diagram corresponding to the experimental measurement. The switch $S_1$ at positions {b} or {c} represents the situation corresponding to the data in panels {b} and {c} respectively. The resistance of the nanowire segment to the contacts N1 and N2 is assumed to be large.}
\label{3}
\end{figure}

\begin{figure}
\begin{center}
\includegraphics[width=6.5in]{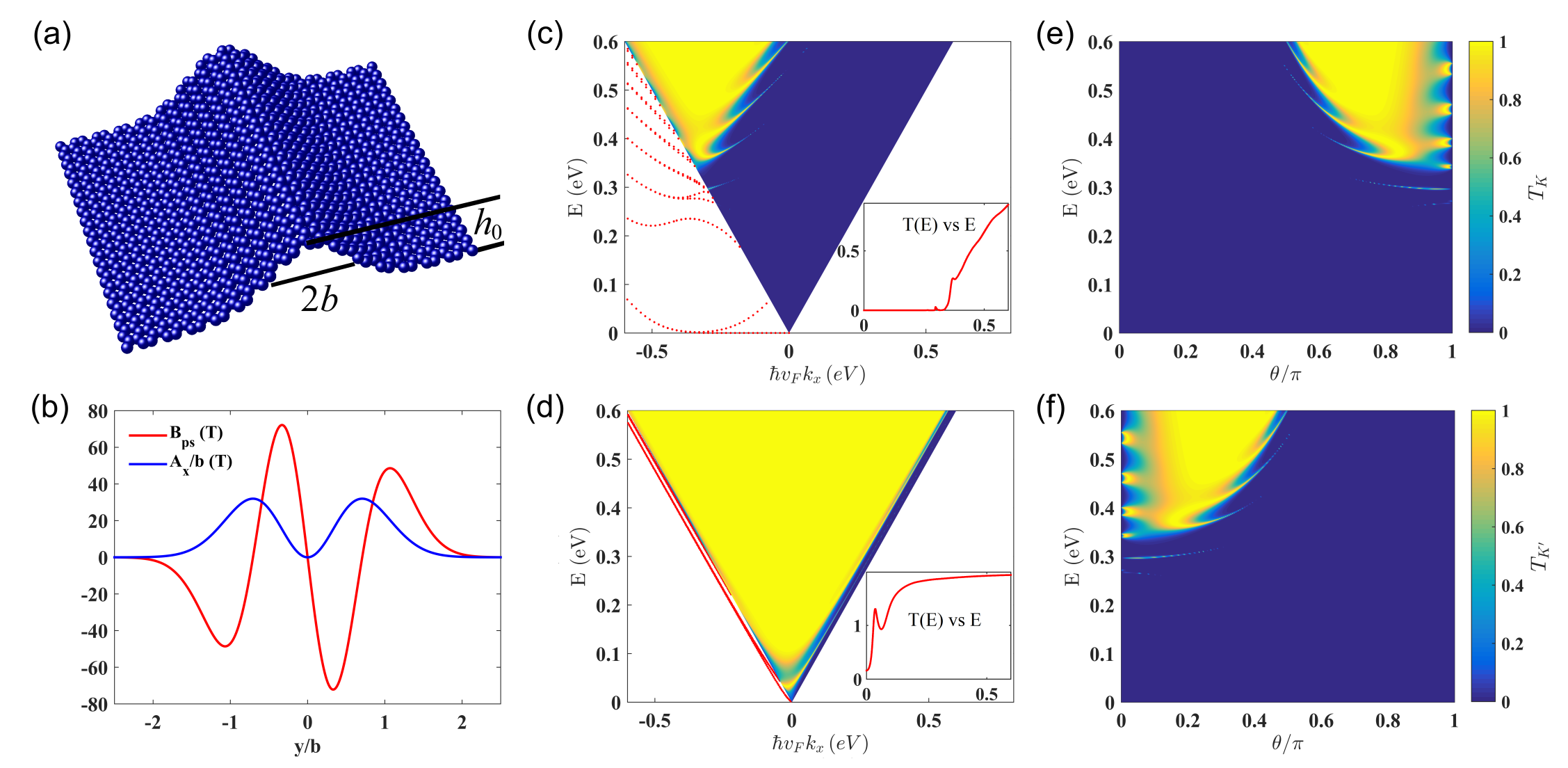}
\end{center}
\caption{{ Theory calculations.} ({a}) Gaussian fold along zigzag direction with height $h_{0}$ and width $b$. ({b}) Pseudo vector potential (blue) and corresponding pseudo-magnetic field (red) as a function of position across the fold region. Fold parameters: $h_0 = 10$nm, $b = 20$nm. ({c}) Transmission probability for K valley states for a Gaussian fold axis oriented along the zigzag direction, as a function of $\hbar v_F k_x$. Red dots indicate bound state dispersion consistent with Coulomb blockade experimental data. Inset: transmission as a function of energy. Axes same as main panel. ({d}) Transmission probability for K valley states for a Gaussian fold axis oriented $1^\circ$ with respect to the armchair direction, consistent with Fabry-Perot oscillations data. Inset: transmission as a function of energy. Axes same as main panel. Transmission coefficient as a function of incident angle for: ({e}) K valley, and ({f}) K' valley. Valley polarization is optimized for incident angles less (greater) than $\theta \sim \pi/2$ with respect to zigzag orientation, i.e., for orientations closer to armchair directions.}
\label{4}
\end{figure}

\end{document}